\documentclass[manuscript]{aastex}

\usepackage{color}
\shorttitle{Meridional Flow Measurements}
\shortauthors{Kholikov et al.}

\begin{document}

\title{MERIDIONAL FLOW IN THE SOLAR CONVECTION ZONE. I. \\
       MEASUREMENTS FROM GONG DATA
           }
\author{S. Kholikov \altaffilmark{1}}
\affil{National Solar Observatories, Tucson, AZ 85719}
\email{kholikov@noao.edu}

\author{A. Serebryanskiy\altaffilmark{2}}
\affil{Ulugh Beg Astronomical Institute, Uzbek Academy of Science, Tashkent 100072, Uzbekistan}

\and

\author{J. Jackiewicz \altaffilmark{3}}
\affil{Department of Astronomy, New Mexico State University, Las Cruces, NM 88003}

\altaffiltext{1}{}
\altaffiltext{2}{}
\altaffiltext{3}{}

\begin{abstract}
Large-scale plasma flows in the Sun's convection zone likely play a major role in solar dynamics on decadal timescales. In particular,   quantifying  meridional  motions is a critical ingredient for  understanding the solar cycle and the transport of magnetic flux. Because the signal of such features can be quite small in deep solar layers and be buried in systematics or noise, the true meridional velocity profile has remained elusive.
We perform time-distance helioseismology measurements on several years worth of GONG Doppler data. A spherical harmonic decomposition technique is applied to a subset of acoustic modes to measure travel-time differences to try to obtain signatures of meridional flows throughout the solar convection zone. Center-to-limb systematics are taken into account in an intuitive, yet ad hoc manner.
Travel-time differences near the surface  that are consistent with  a poleward flow in each hemisphere and are similar to previous work are measured. Additionally, measurements in  deep layers near the base of the convection zone suggest a possible equatorward flow, as well as partial evidence of a sign change in the travel-time differences at mid-convection zone depths.
This analysis on an independent data set using different measurement techniques strengthens recent conclusions that the convection zone may have multiple ``cells'' of meridional flow.  The results may challenge the common understanding of one large conveyor belt operating in the solar convection zone. Further work with helioseismic inversions and a careful study of  systematic effects are needed before firm conclusions of these large-scale flow structures can be made.

\end{abstract}

\keywords{Sun: helioseismology --- Sun: interior --- Sun: oscillations}

\section{Introduction}

Meridional circulation plays a critical  role in models of  solar dynamo, magnetic flux transport, and the solar cycle \citep{glatzmaier1982,wang1989,wang1991,choudhuri1995,dikpati1999,wang2002,nandy2011}. It is well established observationally that meridional flow is poleward in each hemisphere with  an amplitude of about $10-20~{\rm m\,s^{-1}}$ in the near-surface layers, peaking in strength at mid latitudes \citep{duvall1979,hathaway1996,braun1998,hernandez1999,basu1999,hernandez2006,basu2010,hathaway2010,ulrich2010}.

Since mass does not pile up at the poles, it is believed that a return equatorward flow in both hemispheres is operating somewhere in the convection zone, likely near its base. One of the most promising and complete attempts to measure this meridional circulation was during the graduate work of P.~Giles \citep{giles1997,giles2000}.  Using the SOHO spacecraft's Michelson Doppler Imager  (MDI) helioseismic data, Giles found that the poleward meridional flows continued throughout almost all of the convection zone and that there was indirect evidence of a return equatorward flow near the tachocline of a few ${\rm m\,s^{-1}}$. His methods and analysis imposed a constraint of mass conservation. Thus, the  picture that emerged was  of  two closed circulating flows, one cell in each hemisphere, that diverge from the equator at the surface and converge toward the equator in the deep interior.

Since then, other helioseismology studies using a variety of techniques have offered many differing views. For example, \citet{chou2001,beck2002,chou2005} observe an additional ``cell'' of meridional circulation at mid latitudes near the location of the active sunspot latitudes, which is divergent and varies in time. Also, \citet{zhao2004,hernandez2010} found that such a cell has a convergent flow field \citep{cameron2010}. Indeed, large-scale flow profiles (in both meridional and zonal directions) have been found to vary rather strongly with the solar cycle, and several studies have found that the amplitude of the flow  is anti-correlated with the strength of the cycle \citep[e.g.,][]{komm1993,chou2001,haber2002,basu2003,hernandez2008,gizon2008}. The latitudinal extent of the surface poleward flow has  widely varied in the two previous cycles, and some helioseismic measurements indicate a high-latitude, reverse, \emph{equatorward} surface component \citep{dikpati2012}.    To add to the complexity, recent observations have shown an increasing polar flow magnitude as one probes deep into the convection zone \citep{kholikov2012}, and  \citet{hathaway2012} place the equatorward return flow at a depth of 70~Mm.

Recently \citet{zhao2012b}  observed a new systematic center-to-limb signal in time-distance measurements \citep{duvall1993}, which may play a key role in obtaining reliable deep meridional flow measurements and be one of the sources of the discrepant results mentioned above. The approach of \citet{zhao2012b} was to remove the systematic travel-time shifts found in the east-west measurements, after rotation is removed, from the meridional (north-south) measurements. This correction led to consistent helioseismic measurements using several different observables. While the source of this signal is not completely understood, it could be related to existing observational limitations like changes of the line formation heights across the solar disk, which produce additional acoustic travel-time shifts in cross-correlation measurements between different locations. \citet{baldner2012} showed that the effect of the vertical flows from convection in the outer solar convection zone can similarly affect travel-time measurements.


Subsequently, \citet{zhao2013} applied their local techniques to measure two meridional circulation cells in the solar convection zone, while \citet{schad2013} implemented a new global helioseismic analysis that resulted in evidence of a complex multicellular velocity structure. These new and exciting findings  from space-based data present  a potentially revised view of these important large-scale flows.

This paper is the first in a series where we explore meridional circulation using time-distance helioseismology applied to Global Oscillation Network Group (GONG) data.  Here we describe in detail the travel-time measurement procedure we implement, which is non-standard and differs from the methods of \citet{zhao2012,hartlep2013}, for example. We use more than 600 daily sets of GONG velocity images to probe deep into the convection zone. In order to decrease possible geometric and  observational artifacts we have selected dates with a duty cycle of more than 85\% and time periods 
when the solar tilt angle $B_0 \le 4^\circ$. These strict requirements  substantially 
decrease the amount of data that can be used. We assume that the center-to-limb systematic mentioned above is the same in any direction on the solar disk and we compute it only using the equatorial region of the observations.  Travel-time differences are computed for north-south flows and corrected by subtracting the east-west signal. We find strong evidence of a change of sign in the travel-time differences at mid latitudes and depths of about 200~Mm beneath the surface, and compelling evidence that travel times may change sign (thus signaling a flow reversal) at shallower depths of about 50-60~Mm. In Section~\ref{sec:data} we describe the data and analysis procedure, with results and discussion provided in Section~\ref{sec:res}.


\section{Data and analysis technique}
\label{sec:data}

\begin{figure}[t]
  \centering
  \includegraphics[width=.5\textwidth]{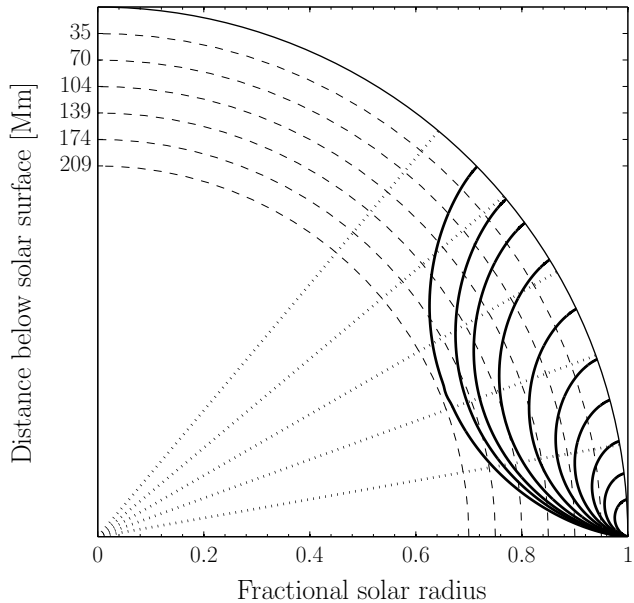}
  \caption{Illustration of approximate ray paths for the 10 phase-speed filters we employ. A realistic solar model is used to trace the paths. We only plot the point-to-arc separation distance for the central distance considered, as listed in Table~\ref{tab:filt}. The five dotted lines are plotted in $10^\circ$ increments. The depth scale in Mm for the lower turning point of the rays is given on the $y$-axis for reference.}
  \label{rays}
\end{figure}

%
%
\begin{table}[t]
  \centering
  \caption{Phase-speed filters used for travel-time measurements in this work.}
  \label{tab:filt}
  \begin{tabular*}{\textwidth}{@{\extracolsep{\fill}} r c c  c c c } 
    \tableline\tableline
         &Min.~$\Delta$&Max.~$\Delta$&$v_{\rm ph}$&$\ell$&Depth\\
    Index&(deg)&(deg)&(${\rm km\,s^{-1}}$)& & ($R_\odot$)\\\hline
    1&2.75&5.75&52.5&250&0.98\\
    2&5.00&8.75&74.3&200&0.96\\
    3&8.00&11.75&96.2&150&0.94\\
    4&12.50&17.75&122.4&100&0.91\\
    5&17.00&23.00&148.7&85&0.88\\
    6&23.00&27.50&183.7&70&0.84\\
    7&28.25&33.50&223.0&60&0.80\\
    8&33.50&38.75&258.0&50&0.77\\
    9&38.00&43.25&284.3&45&0.74\\
    10&42.50&47.00&319.2&40&0.71\\\tableline
  \end{tabular*}
\tablecomments{Depths are approximate and based on ray tracing. Distances for each filter are computed at steps of $0.75$~deg from the minimum $\Delta$ to the maximum $\Delta$. The central phase speed of each Gaussian filter is given by $v_{\rm ph}$,   and the angular degree $\ell$ represents the central value for each filter.}
\end{table}

In this study we utilize GONG  Doppler velocity images, and have selected 652 daily sets of  observations with duty cycle higher than 85\% during the 2004-2012 time period. 
The typical  travel-time measurements are based on cross-correlation (CCF) analysis between two locations on the  solar surface separated by certain distances [$\Delta$] \citep{duvall1993}. It is well known that waves with the same phase speed form a wave packet and propagate along approximately the same 
ray path. In order to increase the signal-to-noise ratio of cross-correlation functions, phase velocity filters are 
used to isolate particular wave packets. To infer the meridional flow signal we measure wave travel times for waves propagating in opposite directions along the same ray path that lies between a pair of points (more precisely, a point and an arc) at constant longitude. In this case, the travel-time difference for waves propagating in the two opposite directions is sensitive only to the meridional (North-South) 
component of the flow \citep{kosovichev1997}. Below we itemize the main steps of the analysis and computation procedure:

\begin{enumerate}

\item Daily velocity time series (1440 images/day) were tracked according to the surface Synodic differential rotation profile \citep{librecht1991} relative to 12:00 (noon)  and remapped into $\sin(\theta) - \phi$ coordinates, where $\theta$ and $\phi$ are latitude and longitude, respectively.

\item Spherical harmonic (SH) decomposition is performed for each image and SH coefficient time series $C_\ell^m$ are obtained for modes $\ell=0-300$ and $m=-\ell,\cdots,\ell$, where $\ell$ is the SH degree and $m$ is the azimuthal order.
  
\item Ten different phase speed filters are employed and designed 
  to cover the approximate depths  of 0.98-0.70R$_\sun$. Details of their parameters are given in Table~\ref{tab:filt}. Only modes within temporal frequencies between 1.8 and 4.5~mHz are retained. The approximate ray paths of the filtered wave packets considered here are illustrated in Fig.~\ref{rays}.
  
\item Phase speed filters are applied to the $C_\ell^m$ time series for each of the 10 cases to yield $\tilde{C}_\ell^m$.
  
\item Velocity images are reconstructed from the  $\tilde{C}_\ell^m$ using the inverse SH decomposition relation:
  \begin{equation}
    Y(\theta,\phi)=\sum_{\ell=0}^{L}\sum_{m=0}^{\ell} \tilde{C}_\ell^m P_\ell^m(\sin\theta)e^{im\phi}.
  \end{equation}
  Here, $P_\ell^m$ are the associated Legendre polynomials, $\theta$ denotes latitude, $\phi$ longitude,  and $L=300$.   The fraction of the reconstructed solar disk was $\pm75\arcdeg$ in latitude and $\pm60\arcdeg$ in longitude relative to the center of the disk. 
  
\item  The CCF between a point and the signal averaged over a 30$\arcdeg$ arc for a given 
  longitude was computed as
  \begin{equation}
    C(\tau,\Delta,\phi)=\int{f(\theta_1,\phi,t)f(\theta_2,\phi,t+\tau)\, {\rm dt}},
  \end{equation}
  where $f$ is the filtered velocity time series, and $\Delta=|\theta_1 - \theta_2|$ is the angular distance between two spatial locations on the solar surface ($\theta_1,\phi$) 
  and ($\theta_2,\phi$). Arcs in the four cardinal directions are considered. For each filtered set of data the cross-correlations were   computed for some range of travel distances with increments of $0.75\arcdeg$ around the maximum of the first bounce in the CCF.  In total  72 correlation functions are computed for travel distances covering $\Delta=2.75\arcdeg-47\arcdeg$ (see Table~\ref{tab:filt}).
 
\item The cross correlations were averaged over about 250 longitude bins in the range  $-45{\arcdeg}\le\phi\le45{\arcdeg}$. Using smaller bands in longitude provides more proper 
center-to-limb corrections, but leads to a decrease in the signal-to-noise ratio of cross correlation measurements. Simple comparison  using narrower bands showed significant increase of the variance of individual measurements. Since we are interested in travel time differences of about 1 s we decided to use a wider longitude range. 
  
\item Northward and southward travel times were obtained by fitting a Gabor wavelet function to both the positive and negative lags ($\tau$) of the cross correlations as
  \begin{equation}
    G(\tau)=A\,\exp\left(-\frac{(\tau-\tau_{\rm g})^2}{2\,\sigma^2}\right)\cos(\omega_{0}(\tau-\tau_{\rm ph})),
  \end{equation}
  where parameters $A$, $\tau_{\rm g}$, $\tau_{\rm ph}$, $\omega_{0}$ and $\sigma$
  are the amplitude, group and phase travel times, mean frequency, and width of wave packet, respectively.
 
\item The difference between two oppositely directed travel times is computed for travel
  distances $\Delta$ corresponding to each phase velocity filter. We use the convention of ``south minus north'' (SN) travel-time differences.

\end{enumerate}


In addition, travel-time differences in the ``east minus west'' (EW) direction have also been  computed using all of the exact steps of the data processing procedure outlined above. The travel times for these measurements were  averaged over $\pm20\arcdeg$ in latitude. A constant shift due to internal solar rotation is observed and removed for each travel distance measurement. These measurements are used to correct the systematics for the meridional observations.

\section{Results and Discussion}
\label{sec:res}

\begin{figure}
  \centerline{
    \includegraphics[width=.5\textwidth,clip=]{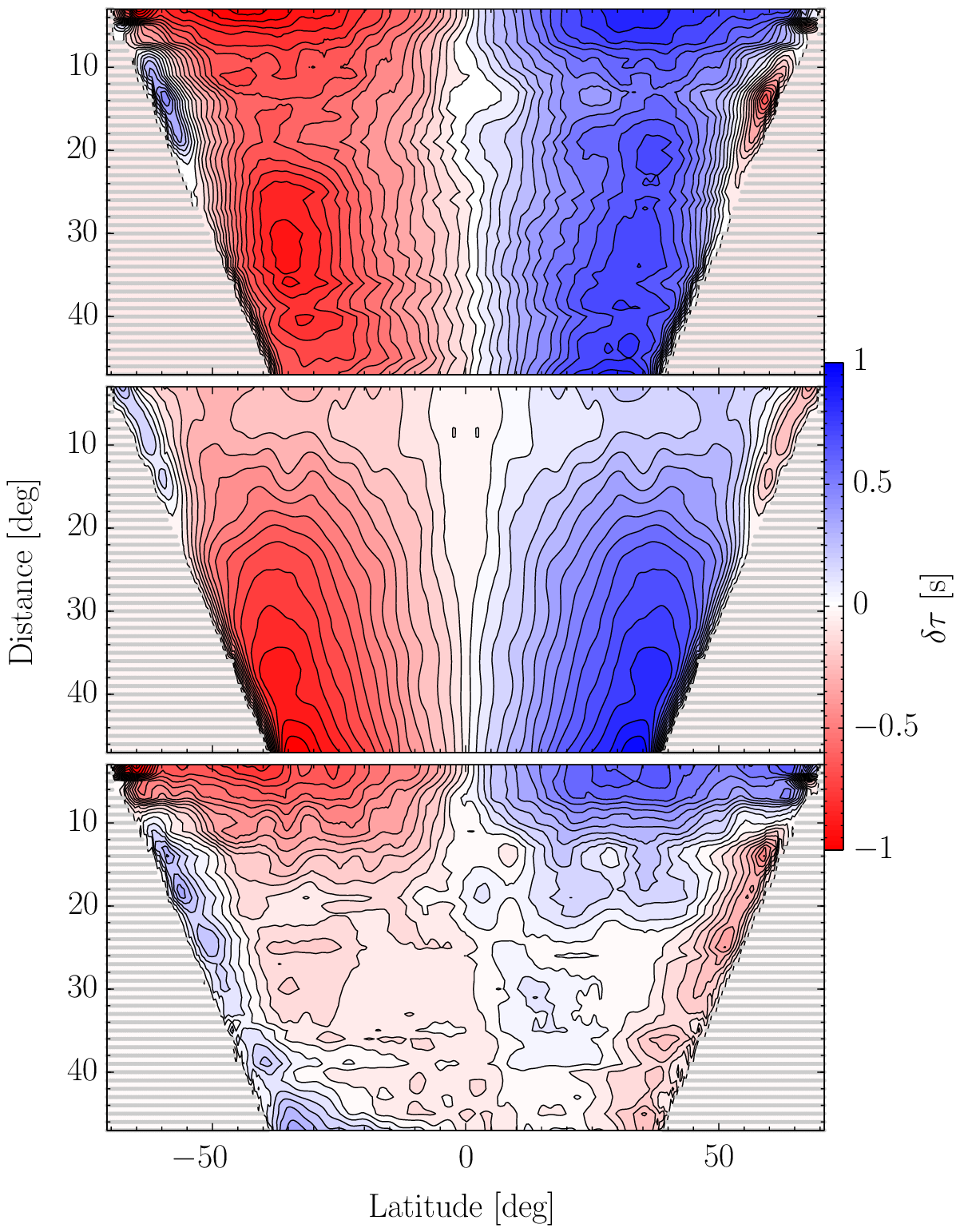}
    \includegraphics[width=.5\textwidth,clip=]{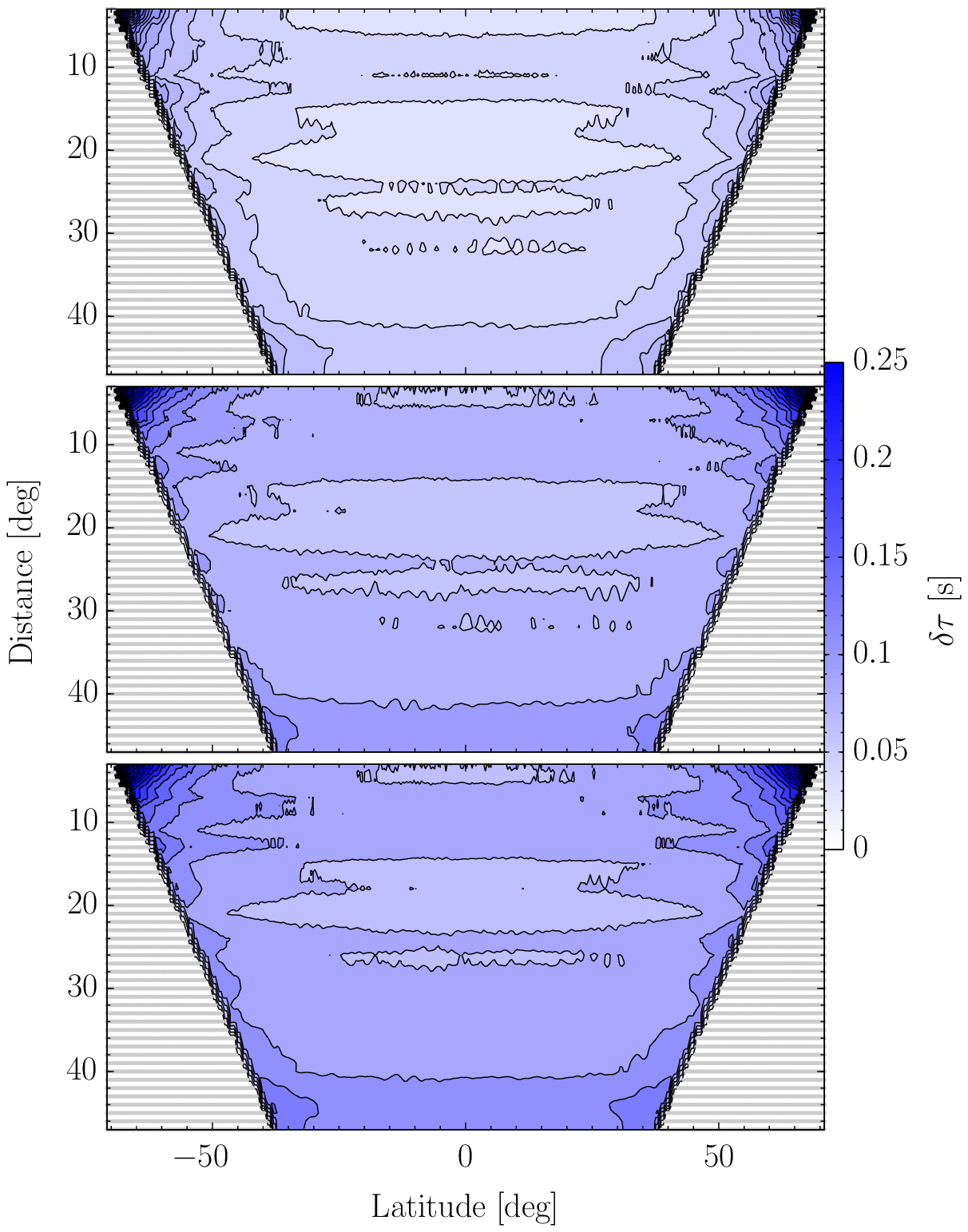}} 
  \vspace{-.4\textheight}
  \centerline{\hspace{.06\textwidth}\bf\Large\color{black}{(a)}\hspace{.45\textwidth}\bf\color{black}{(b)}\hfill}
  \vspace{.35\textheight}
  \caption{Travel-time difference  maps  obtained using 652 daily sets of Doppler velocity images. Column (a) shows the SN, EW, and SN-EW contour maps from top to bottom, respectively. Column (b) plots the corresponding measurement uncertainties associated with each panel in column (a). Note the $x$ axis in the middle panel in each column is the longitude, with the same numerical scale values as shown for latitude ($\pm 75\arcdeg$). Hatched regions show where no measurements were computed due to limb constraints.\label{maps}}
\end{figure}

\begin{figure}
  \centerline{
    \includegraphics[width=.5\textwidth,clip=]{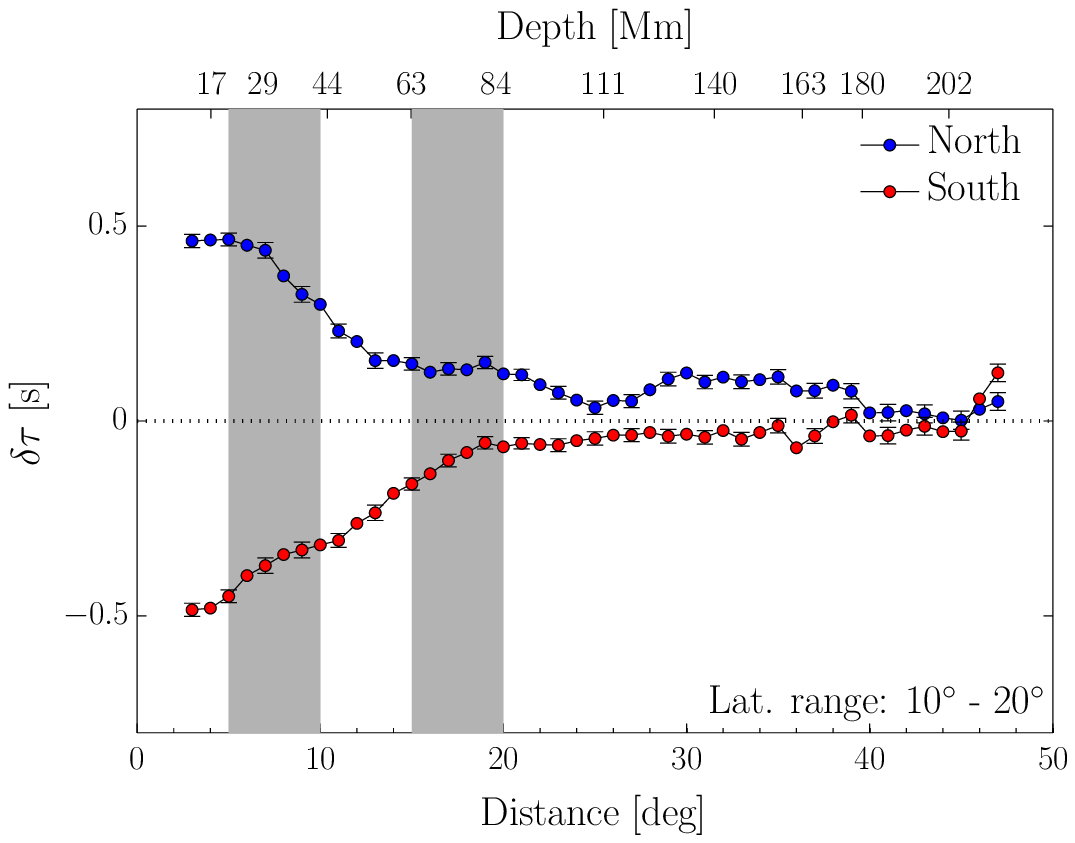}
    \includegraphics[width=.5\textwidth,clip=]{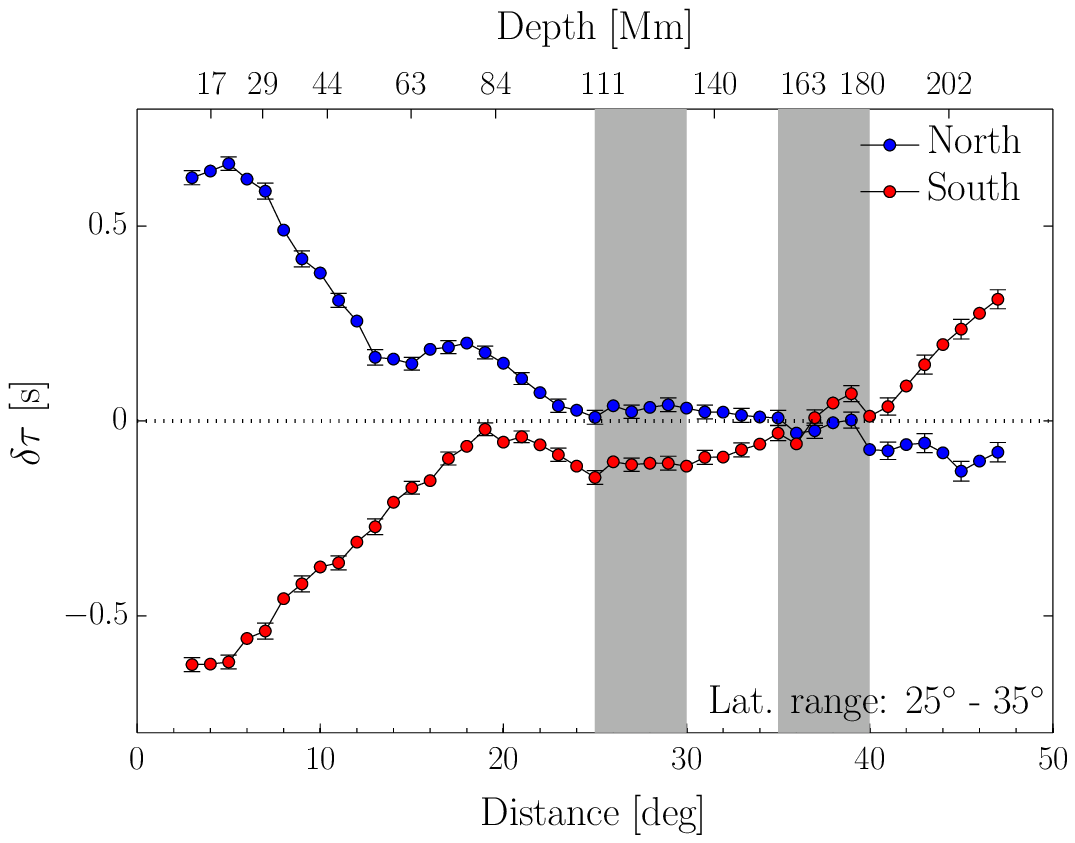}} 
  \vspace{-.2\textheight}
  \centerline{\hspace{.07\textwidth}\bf\large\color{black}{(a)}\hspace{.47\textwidth}\bf\color{black}{(b)}\hfill}
  \vspace{.17\textheight}
  \centerline{
    \includegraphics[width=.5\textwidth,clip=]{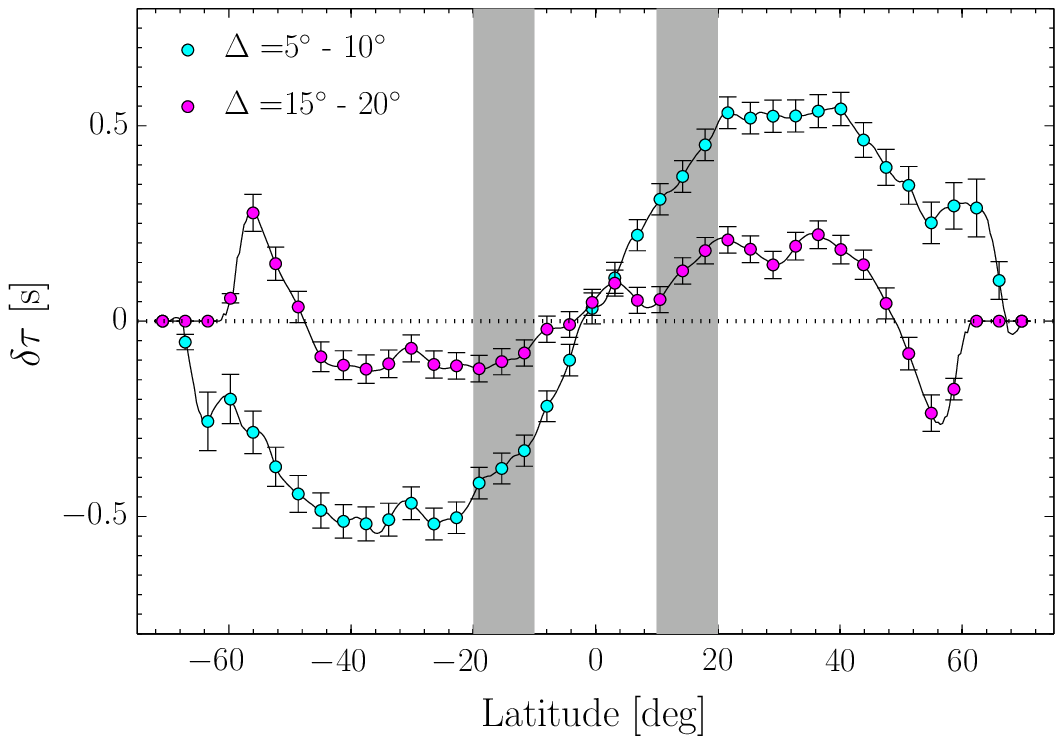}
    \includegraphics[width=.5\textwidth,clip=]{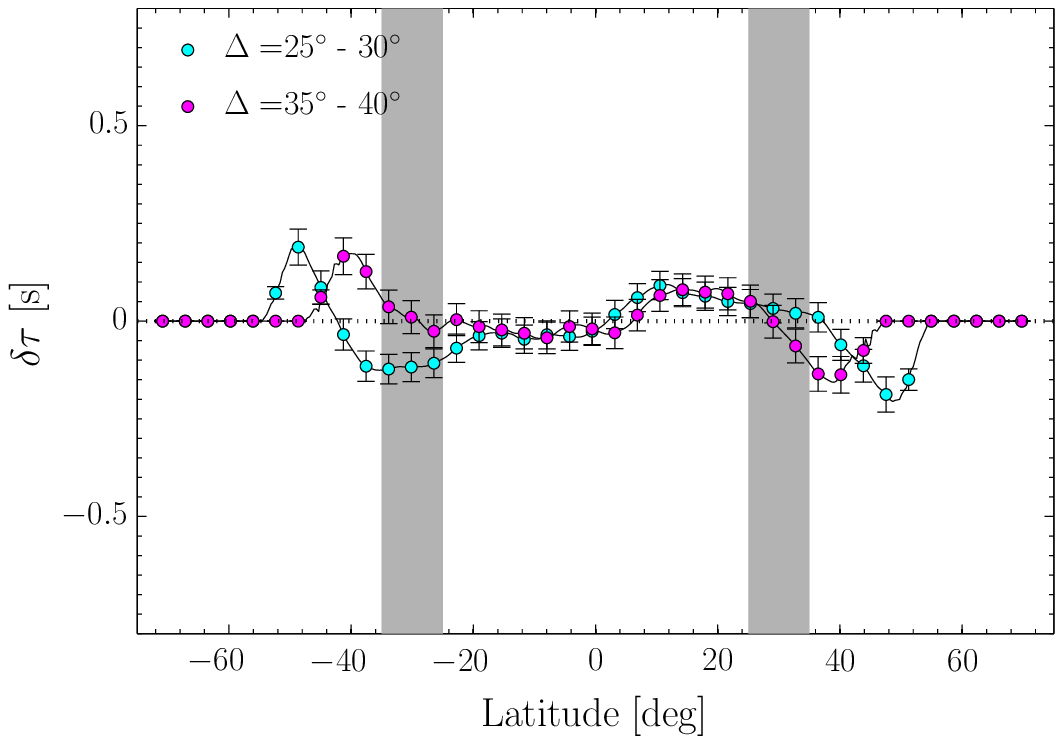}}
  \vspace{-.1\textheight}
  \centerline{\hspace{.07\textwidth}\bf\large\color{black}{(c)}\hspace{.47\textwidth}\bf\color{black}{(d)}\hfill}
  \vspace{.05\textheight}
  \caption{Cuts through depth and latitude of corrected SN travel-time differences. The top row panels (a) and (b) show the travel-time differences in each hemisphere as a function of measurement distance ($\Delta$) for the latitude range averaged over  the  $10^\circ$ band  noted in the figure. A proxy for the lower turning point depth for each travel distance is shown on the upper $x$ axis. Highlighted in gray are the travel distances shown in the corresponding plots below. Panels (c) and (d) shows travel times as a function of latitude averaged over an interval in distances of $5^\circ$. The latitude ranges  in panels (a) and (b) are given in the gray boxes of panels (c) and (d). The  uncertainties are shown for all cases and are plotted only at staggered data points for clarity.\label{cuts}}
\end{figure}

The top left panel of Fig.~\ref{maps} shows an average over 652 days of SN travel-time differences  
presented as a function of latitude and travel distance. Each point at a given travel distance corresponds to the middle 
position between a point and an arc in our cross-correlation scheme. To avoid very high latitude information 
where the endpoints of the cross correlations lie, the measurements are cut off as a function of distance. The uncertainties are given in the second column, computed from the dispersion in individual measurements for each longitude and each day. These are typically a very small percentage of the averaged signal.

Signatures of poleward meridional flow in each hemisphere are clearly seen in Fig.~\ref{maps}(a). The color convention in this figure is such that blue is consistent with a flow toward the North Pole, and red a flow toward the South Pole. Indeed, in addition to a peak at mid latitudes as expected, an increase in the travel-time difference with depth (i.e., travel distance) is also observed. We expect this to be due to one or several systematics. To explore this further, EW travel times computed from the same dataset are shown in the middle panel of Fig.~\ref{maps} as a function of longitude on the $x$ axis. The EW map has been symmetrized about the central meridian, as we expect there to be no significant differences between the two (east/west) hemispheres since the data have been tracked to account for differential rotation. These measurements show a similar pattern of center-to-limb variation as the SN map.

\citet{zhao2012} reported a very detailed analysis of travel-time measurements from different observables. Since they found that the shape and magnitude of center-to-limb variations  is quite different for Doppler, continuum, line core and line depth of HMI measurements, one might conclude  that these variations are not caused by any large-scale sub-surface flow of solar origin. Here we follow the same procedure and ``correct'' the SN measurements by subtraction of the EW measurements, the result shown in the  bottom panel of  Fig.~\ref{maps}(a). This correction removes the tendency of the travel times to increase with depth. Furthermore, some evidence of  sign changes  can be seen.

Figure~\ref{cuts} shows various cuts through the travel-time difference maps. Panels (a) and (c) are cuts at lower latitudes and shorter travel distances, while panels (b) and (d) are for mid latitudes and larger travel distances. These figures confirm that travel-time differences are strongest at mid latitudes around $30^\circ$ for a range of depths, as has been observed in past studies. This representation shows a clear yet peculiar asymmetry between the northern and southern hemispheres. Most importantly, we also observe evidence that a change in sign occurs in the measurements for two cases: (1) at high latitudes in each hemisphere for travel distances greater than about $15^\circ$; and (2) for large distances for most latitudes greater than about $20^\circ$ in each hemisphere.

Indeed, if large-scale flows are responsible for these signals, Figs.~\ref{cuts}(a)-(b)  show a tendency for the flow to approach a change of sign at skip distances of $15^\circ-20^\circ$ for a broad latitude range. At larger distances this signal then resurrects its poleward sense, eventually reversing again at the deepest probe depths. This very broadly suggests a multicellular structure as discussed in \citet{zhao2012} and \citet{zhao2013}, who found poleward flows down to $0.91~R_\odot$, equatorward flows in the $0.82 - 0.91~R_\odot$ range, and then poleward again beneath that.  Very recent work by  \citet{schad2013} reports yet another measurement of multicellular structure of the meridional flow using a different, global approach.

We caution that the change in sign at all distances at the maximal latitudes considered here (most evident in Fig.~\ref{maps}) could be due to  a systematic caused by the solar $B_0$ variation, as demonstrated  recently in \citet{kholikov2013}. However, in the measurements here such an artifact is somewhat puzzling since we have restricted the data coverage to epochs when this angle is small.   Another possible cause could simply be the use  of the ad hoc correction method and any of its inherent systematics.  Also evident in the measurements is a north-south hemispheric asymmetry from GONG and space-based data that has been noted in previous works  \citep[e.g.,][]{zaatri2006,rightmire2012}.

The real origin of the center-to-limb variation across the solar disk is not well understood at present. The work of \citet{baldner2012} in explaining it is promising. Anomalous artifacts were even identified as early as \citet{duvall2009}, who consider effects due to the finite speed of light in meridional-flow measurements as causing an overall inflow towards the disk center. While only several seconds, the proper correction  for this effect individually actually tends to add to the already unphysical increasing travel-time difference signal with depth as observed in the top panel of Fig.~\ref{maps}(a). Presumably this systematic is already accounted for in the east-west subtraction correction implemented here, although more confidence in such an approach is certainly needed and is the focus of current work.


Nevertheless, we find strong evidence of variations in depth of the large-scale flow in the solar convection zone. The deepest measurements where the ray path is horizontal and less sensitive to surface flows show strong evidence of a change of sign. The tendency for the travel-time shifts to approach zero at mid-convection zone depths is also intriguing, as one must recall that these measurements are integrated over depth and smoothed to some degree. Inversions may separate the two directional components of the flow and provide amplitudes and a more accurate depth structure.

We have shown robust travel-time measurements of the meridional flow signature in the solar convection zone using GONG data and independent measurement techniques. Preliminary evidence of a change of sign, indicating an equatorward return flow at one or several depths in the convection zone is observed, and is approximately consistent with results found in other recently published work. Overall the findings might suggest multicellular structure in the large-scale flows in the Sun. Only a consistent inversion procedure and a very careful treatment of the systematics can unravel the significance of these trends in the measurements. A forthcoming paper will show such inversions and discuss the implications for convection-zone dynamics.

\acknowledgments
We thank an anonymous referee for helpful comments. A.S. is supported by a grant of the Uzbek Academy of Sciences FA-F02-F028. J.J. acknowledges support from NASA contract \#NNX09AP76A and NSF AST-0849986. S.K. was supported by NASA grant NNX11AQ57G to the National Solar Observatory.
This work utilizes GONG data obtained by the NSO Integrated Synoptic Program (NISP), managed by the National Solar Observatory, which is operated by AURA, Inc. under a cooperative agreement with the National Science Foundation. The data were acquired by instruments operated by the Big Bear Solar Observatory, High Altitude Observatory, Learmonth Solar Observatory, Udaipur Solar Observatory, Instituto de Astrof\'isica de Canarias, and Cerro Tololo Inter-American Observatory.

 

\end{document}